\documentclass[apjl]{emulateapj}
\shorttitle{Boron Synthesis in SNe Ic}
\shortauthors{Nakamura et al.}

\def\Msun{~M_{\odot} }

\begin{document}

\title{
Boron Synthesis in Type Ic Supernovae
}

\author{Ko Nakamura$^{1}$, Takashi Yoshida$^{2}$, Toshikazu Shigeyama$^{2,3}$ and Toshitaka Kajino$^{1,2}$}
\affil{$^1$National Astronomical Observatory of Japan, Mitaka, Tokyo, Japan\\
        $^{2}$Department of Astronomy, Graduate School of Science, 
        University of Tokyo, Tokyo, Japan\\
        $^{3}$Research Center for the Early Universe, Graduate School of Science, 
        University of Tokyo, Tokyo, Japan\\
        }

\begin{abstract}
We investigate the $\nu$-process in an energetic Type Ic supernova (SN Ic) and resultant productions of the light elements including boron and its stable isotopes.
SN Ic is a very unique boron source because it can produce boron through not only spallation reactions as discussed in \citet{Nakamura04} but also the $\nu$-process.
The $\nu$-process is considered to occur in core-collapse supernovae and previous studies were limited to Type II supernovae (SNe II).
Although the progenitor star of an SN Ic does not posses a He envelope so that $^7$Li production via the $\nu$-process is unlikely, $^{11}$B can be produced in the C-rich layers.
We demonstrate a hydrodynamic simulation of SN Ic explosion and estimate the amounts of the light elements produced via the $\nu$-process for the first time, and also the subsequent spallation reactions 
between the outermost layers of compact SN Ic progenitor and the ambient medium.
We find that the $\nu$-process in the current SN Ic model produces a significant amount of $^{11}$B, 
which is diluted by $^{10}$B from spallation reactions to get closer to B isotopic ratios observed in meteorites.
We also confirm that high-temperature $\mu$ and $\tau$ neutrinos and their anti-neutrinos, 
reasonably suggested from the compact structure of SN Ic progenitors, 
enhance the light element production through the neutral-current reactions, which may imply an important role of SNe Ic in the Galactic chemical evolution.\end{abstract}

\keywords{nuclear reactions, nucleosynthesis, abundances --- supernovae: general --- stars: abundances}

\section{Introduction}\label{sec-int}
Stars with mass more than about $10 \Msun $ are known to culminate their evolution as supernova explosion. Most of the gravitational energy released from formation of proto-neutron star is carried away by neutrinos and total number of emitted neutrinos amounts to more than $10^{58}$. Although cross sections of neutrino-nucleus reactions are very small, a fraction of such a huge number of neutrinos, in particular $\mu$- and $\tau$- neutrinos with higher temperatures than electron neutrinos, can scatter off nuclei to the excited states, induce ejection of neutron, proton, or $\alpha$-particle, and leave observational signature of the $\nu$-process nucleosynthesis in the yields of some species \citep{Woosley90}. 

Massive stars are one of the astrophysical sites for the production of light elements via the $\nu$-process, mainly contributing to the production of $^7$Li and $^{11}$B \citep[e.g.][]{Woosley90,Heger05,Yoshida04}. $^7$Li is mainly produced in He-rich envelopes and $^{11}$B in inner C-rich shells. The stellar materials including newly-synthesized light elements are accelerated by strong shock wave, a part of which is destroyed by shock heating, and finally ejected into the interstellar matter to make a partial contribution to the next generations of star formation.

Light element isotopes are also known to be produced from some other processes; $^7$Li from big bang nucleosynthesis, $^7$Li and $^{11}$B from AGB stars and novae \citep{Cameron55}, and all stable isotopes of Li, Be, and B (LiBeB) from Galactic cosmic-ray (CR) nucleosynthesis \citep[e.g.][]{Meneguzzi71, Suzuki02, Rollinde08}.

Despite a great deal of efforts dedicated to investigate these processes, 
the origins of LiBeB are not fully understood and there remain some observational features which conflict with theoretical predictions.
For instance, theoretical predictions for LiBeB production from Galactic CRs alone indicate quadratic relation between BeB abundances and metallicity, while observations clearly show linear relation.
CNO CRs from superbubbles \citep{Higdon98}, supernovae or Wolf-Rayet stars \citep{Ramaty97,Yoshii97}, and Type Ic supernovae \citep{Fields02, Nakamura04} have been suggested to account for the observed linearity.
However, the CR spallations cannot be the only source of boron isotopes because the CR spallations do not reproduce the high $^{11}$B-to-$^{10}$B abundance ratio as observed in meteorites, and another source of $^{11}$B such as supernova $\nu$-process is necessary.

It is required to accurately and quantitatively evaluate each contribution of the LiBeB production process in order to solve these problems.
In this Letter we 
report that energetic core-collapse Type Ic SNe (SNe Ic) could be a viable astrophysical site where light elements including boron isotopes are produced.
Our nucleosynthesis model takes account of the $\nu$-process in SNe Ic for the first time and also spallation reactions of accelerated SN ejecta interacting with interstellar/circumstellar matter, which do not conflict with the observed linear relation between BeB abundances and the metallicity in metal-deficient halo stars.
The progenitor of an SN Ic is a C/O star and its H and He envelopes have been stripped
during the stellar evolution.
Although the explosion mechanism of SNe Ic has not been clarified,
neutrinos should be one of the main carriers of the gravitational
energy release from the collapsing core.
The robustness of neutrino emission from a collapsing proto-neutron star which was evolved from a $\sim 40 M_\odot$ progenitor star was studied in numerical simulation \citep{Sumiyoshi06,Sumiyoshi07}.
Even in the case that the central core of such a star eventually turns into a black hole, a temporally formed proto-neutron star is shown to emit a huge amount of neutrinos before collapsing to the black hole.
Even after the black hole formation, neutrinos are being emitted from accretion disk 
\citep[e.g.][]{Kohri05,Surman05}.
Therefore, a huge amount of neutrinos are emitted from the central region
of an SN Ic so that the $\nu$-process can operate in the exploding supernova material.
In \S 2, we present our model of supernova explosion (\S \ref{sec-sn}), the $\nu$-process nucleosynthesis (\S \ref{sec-np}), and the spallation reactions (\S \ref{sec-sp}).
We discuss our calculated results in \S 3 where we 
clarify the important role of SNe Ic in the light element synthesis.

\section{Calculations and Results}\label{sec-clc}
\subsection{\textmd{\textit{Supernova Explosion}}}\label{sec-sn}
We consider a very energetic explosion of a 15 $M_{\odot}$ C/O star with the explosion energy $E_{\rm ex}=3 \times 10^{52}$ ergs corresponding to SN 1998bw \citep{Nakamura01}.
The explosion energy is released at the center of the progenitor star as thermal energy.
Resulting shock wave accelerates the stellar materials and explodes the progenitor as a supernova.
Time evolutions of physical quantities in the progenitor are calculated with 1D hydrodynamic code which takes account of the effects of special relativity.
We solve the special relativistic hydrodynamic equations in Lagrangian coordinates with an ideal equation of state
involving gas and radiation pressure. 
Adiabatic indices are treated as functions of pressure and gas density. 
Details on the numerical method are described in \citet{Nakamura04}.

\subsection{\textmd{\textit{The Neutrino-Process}}}\label{sec-np}
Light element synthesis in the Type Ic supernova is calculated as a post process.
We use a nuclear reaction network consisting of 291 species of nuclei
and taking account of the $\nu$-process \citep{Yoshida08}.

Neutrinos are emitted from a collapsing proto-neutron star 
\citep[e.g.][]{Sumiyoshi06} and/or the innermost region just above 
a black hole \citep[e.g.][]{Surman05}.
Neutrino properties,
particularly in SNe Ic, are still uncertain.
We therefore use a supernova-neutrino model \citep{Yoshida08} 
that the neutrino luminosity decreases exponentially with a time scale of 3 s and that the neutrino temperature of each species does not change with time.
The total energy carried out by neutrinos
is assumed to be $3 \times 10^{53}$ ergs.

We consider two different models for neutrino temperatures.
For both models the temperatures of $\nu_e$ and $\bar{\nu}_e$ are set to be $T_{\nu_e}$ = 3.2 MeV and $T_{\bar{\nu}_e}$ = 5 MeV.
In the first model, 
the temperature of the other flavors, $\nu_{\mu,\tau}$ and $\bar{\nu}_{\mu,\tau}$, takes an identical value of $T_{\nu_{\mu,\tau}}$=6 MeV.
This temperature model is referred to as
the ^^ ^^ standard'' $T_{\nu_{\mu,\tau}}$ model.
The light element synthesis in an $\sim 20 M_\odot$ 
Type II supernova with the standard temperature model well reproduces the supernova contribution of the $^{11}$B production during Galactic chemical evolution 
\citep{Yoshida05,Yoshida08}.
The third peak of r-process elements is also reproduced well in neutrino-driven
wind models using these temperatures \citep{Yoshida04}.
In the second model, the temperature of $\nu_{\mu,\tau}$ and 
$\bar{\nu}_{\mu,\tau}$ is set to be $T_{\nu_{\mu,\tau}}$ = 8 MeV.
This temperature model is referred to as the
^^ ^^ high'' $T_{\nu_{\mu,\tau}}$ model.
A newly forming proto-neutron star in an SN Ic should be more compact than that of an SN II.
Therefore, the neutrino temperature of the collapsing proto-neutron star 
is larger than that of SN II \citep[e.g.][]{Sumiyoshi06}.

\begin{figure}
        \includegraphics[width=12cm,angle=270]{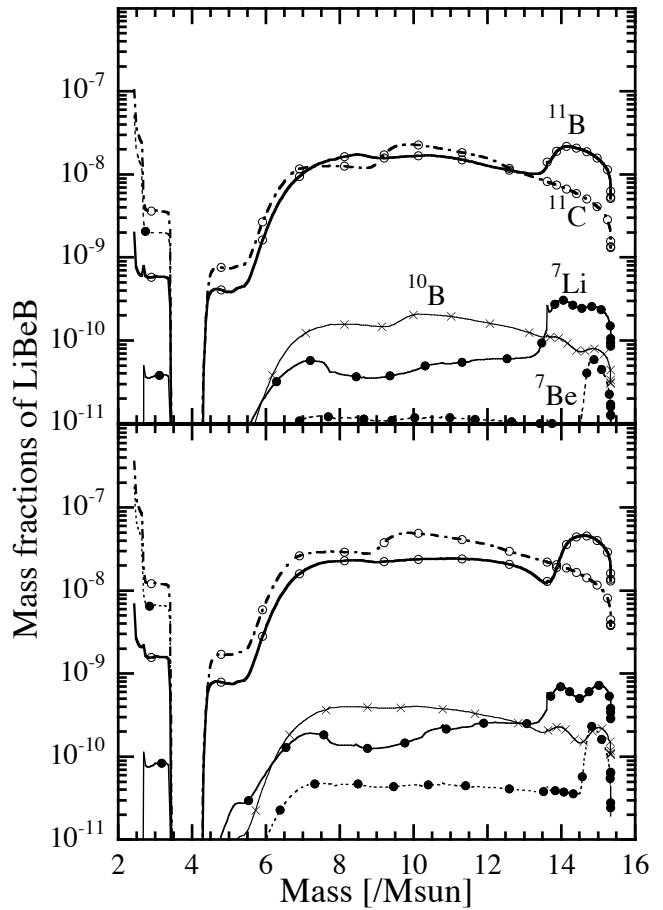}
	\caption{The mass fraction distributions of light elements produced 
via the $\nu$-process as functions of the mass coordinate. 
Shown are the cases of the SN 1998bw model 
($M_{\rm CO} = 15 M_{\odot}$, $E_{\rm ex} = 3 \times 10^{52}$ ergs) with 
$T_{{\nu}_e} = 3.2$ MeV, $T_{{\bar{\nu}}_e} = 5$ MeV, and  
$T_{\nu_{\mu \tau}} = 6$ MeV ({\it top}) and $8$ MeV ({\it bottom}) at 75 seconds after the energy release.}
\label{fig-np}
\end{figure}

Calculated mass fraction distributions of $^{11}$B, $^{11}$C, $^{10}$B, $^7$Li,
and $^7$Be in the SN Ic model are shown in Figure 1.
The yield of each species strongly depends on the 
branching ratios of the decay channel of $^{12}$C-neutrino reactions.
The light elements are mainly produced in the O/Ne layer
($ M_r \ga 7.4 M_\odot$).
The mass fractions in the O/Si layer 
($4.6 M_\odot \la M_r \la 7.4 M_\odot$)
are smaller than in the outer layer.
The Si/S layer ($3.7 M_\odot \la M_r \la 4.6 M_\odot$) where carbon completely burned out
presents no light-element production.
The mass fractions of some light elements are also large in the innermost
region where $\alpha$-rich freeze out 
is
achieved ($M_r \la 2.6 M_\odot$)
and complete Si-burning occurs ($2.6 M_\odot \la M_r \la 3.5 M_\odot$).
There is no qualitative difference in the mass fraction distributions between the standard and high $T_{\nu_{\mu,\tau}}$ models
except that the mass fraction of each species in the high $T_{\nu_{\mu,\tau}}$ model at a given mass coordinate is roughly twice as large as the one in the standard $T_{\nu_{\mu,\tau}}$ model.

Several light elements produced from $^{12}$C-neutrino reactions are exposed to explosive nucleosynthesis when the shock arrives.
$^{11}$B, $^{11}$C, and $^{10}$B are destroyed subsequently by collisions with protons and $\alpha$-particles.
$^7$Li and $^7$Be are photodisintegrated to $^3$H and $^3$He, respectively.
After the shock passage, the exploding materials are still being irradiated by neutrinos so that light elements are produced through 
the $\nu$-process again.
This is the reason why larger mass fractions of $^{11}$B, $^7$Li, and $^7$Be in the outermost region ($M_r \ga 13.5 M_\odot$) 
survive, being avoided from explosive nucleosynthesis.

In the innermost region, light elements are produced after the termination
of the nuclear statistical equilibrium.
The main product in $\alpha$-rich freeze out is $^4$He.
About 20 \% of $^4$He by mass fraction is also produced in complete Si-burning.
The $\nu$-process of $^4$He produces $^3$H and $^3$He in cooling materials, followed by $\alpha$-capture reactions to produce $^7$Li and $^7$Be.
Furthermore, $^{11}$B and $^{11}$C are produced by $\alpha$-captures,
and $^{10}$B by $^7$Be($\alpha,p)^{10}$B reaction.

Table \ref{tbl-yield} summarizes the total yields of the light elements produced through the $\nu$-process.
In both models of  $T_{\nu_{\mu,\tau}}$ significant amount of $^{11}$B of order $10^{-7} M_\odot$ is synthesized.
It should be noted that the amounts of $^{11}$B include $^{11}$C which decays to $^{11}$B in 20 minutes.
The yield of $^7$Li is of the order of $10^{-9} - 10^{-8} M_\odot$ 
in the SN Ic model.
This yield is much smaller than that produced in SNe II 
\citep[e.g.][]{Yoshida08}. 
Most of $^7$Li in SNe II is produced in the He-rich layer.
In contrast, almost all H and He layers of SN Ic progenitors have been
stripped via stellar wind and/or binary effect before explosion.
The yields of $^6$Li, $^9$Be, and $^{10}$B are of the order of
or below $10^{-9} M_\odot$ due to smaller branching ratios of neutrino-$^{12}$C reactions
\citep{Yoshida08}.

\begin{deluxetable}{lccc}
\tablewidth{0pt}
\tablecaption{Light Element Yields from SN Ic Model}
\tablehead{
\colhead{Species} & \colhead{Standard $T_{\nu_{\mu,\tau}}$} & 
\colhead{High $T_{\nu_{\mu,\tau}}$} & 
\colhead{Spallation} \\
\colhead{} & \colhead{($M_\odot$)} & \colhead{($M_\odot$)} & 
\colhead{($M_\odot$)}
}
\startdata
$^6$Li & $1.67 \times 10^{-11}$ & $5.91 \times 10^{-11}$ &
$2.38 \times 10^{-7}$ \\
$^7$Li & $7.41 \times 10^{-9}$ & $2.50 \times 10^{-8}$ &
$3.31 \times 10^{-7}$ \\
$^9$Be & $4.49 \times 10^{-11}$ & $1.08 \times 10^{-10}$ &
$9.99 \times 10^{-8}$ \\
$^{10}$B & $1.29 \times 10^{-9}$ & $2.78 \times 10^{-9}$ &
$4.38 \times 10^{-7}$ \\
$^{11}$B & $2.69 \times 10^{-7}$ & $5.46 \times 10^{-7}$ &
$1.34 \times 10^{-6}$
\enddata
\label{tbl-yield}
\end{deluxetable}

\subsection{\textmd{\textit{Spallation Reactions}}}\label{sec-sp}
Supernova explosions accelerate and expel the stellar materials into the interstellar space, which pollutes the universe with newly synthesized elements.
The progenitors of SNe Ic are so compact that a small fraction of ejecta can be accelerated nearly to the speed of light.
Thus accelerated surface layers of SNe Ic, which are composed of C and O, interact with interstellar matter \citep{Nakamura04} or circumstellar matter \citep{Nakamura06}, and produce the light element isotopes via spallation reactions of CNO with protons or $\alpha$-particles.

Here we use the model for SN 1998bw
constructed by \citet{Nakamura04}.
About $0.3 \%$ ($0.04 M_\odot$) of the ejecta attain enough energy ($\gtrsim 10$ MeV/A) to undergo spallations (see eq. \ref{eq-ej}). 
We solve the transfer equation for each element $i$ expressed as
\begin{equation}\label{eqn-tfr}
		\frac{\partial F_i(\epsilon,t)}{\partial t} = \frac{\partial [\omega_i (\epsilon) F_i(\epsilon,t)]}{\partial \epsilon}
		  - \frac{F_i(\epsilon,t)}{\Lambda}  \rho v_i(\epsilon),
\end{equation}
where $\Lambda$ is the loss length in g $\mathrm{cm}^{-2}$, $\rho$ denotes the mass density of the ISM, $v_i(\epsilon)$ the velocity of the element $i$ with an energy per nucleon of $\epsilon$.
The initial condition for the mass of element $i$ with an energy per nucleon $\epsilon$ at time $t=0$, $F_i (\epsilon,t=0)$, is derived from the numerical calculations of explosions or an empirical formula 
\begin{equation}\label{eq-ej}
\frac{M(>\epsilon)}{M_{\rm ej}}= 1.9 \times 10^{-4} \left(\frac{E_{\rm ex}/10^{51}\,\mbox{ergs}}{M_{\rm ej}/1\, \Msun}\right)^{3.4} \left(\frac{\epsilon}{10\,\mbox{MeV}}\right)^{-3.6},
\end{equation}
where $M_{\rm ej}$ is the 	mass of the ejecta. 
The yield of a light element $l$ via the $i+j \rightarrow l+\cdots$ reaction is estimated from
\begin{equation}
	\frac{dN_l}{dt} = n_j \int \sigma_{i,j}^l(\epsilon) \frac{F_i(\epsilon,t)}{A_i m_p} v_i(\epsilon) d\epsilon,
\end{equation}
where $N_l$ is the number of the produced light element $l$, $n_j$ is 
the number density of an element $j$ in the interstellar matter, 
$\sigma_{i,j}^l$ the cross section of $i+j \rightarrow l+\cdots$ reaction given by \citet{Read84}, and $A_i$ the mass number of the element $i$.
The interstellar matter is assumed to consist of neutral H and He with the number densities of  $n_{\rm H}=1\,\mathrm{cm}^{-3}$ and $n_\mathrm{He} = 0.1\,\mathrm{cm}^{-3}$.

Calculated results show that the mass of $^{11}$B produced via spallation reactions is $1.3 \times 10^{-6} M_{\odot}$, which is larger than that synthesized via the $\nu$-process even in the high $T_{\nu_{\mu,\tau}}$ model (Table \ref{tbl-yield}).
The resultant isotopic ratio of $^{11}$B/$^{10}$B ($\sim 3$) from spallation reactions in this model is predominantly determined 
by the ratio of cross sections of the reaction p,$\alpha$ + O $\rightarrow ^{10}$B to that of p,$\alpha$ + O $\rightarrow ^{11}$B.
The yields of the other light elements are listed in Table \ref{tbl-yield}.

\section{Conclusions and Discussion}
\citet{Nakamura04} concluded that SNe Ic may not play an important role in B isotope productions because of the low isotope ratios $^{11}{\rm B}/^{10}{\rm B} \sim 2.8$ compared with observations in meteorites ($4.05 \pm 0.05$). 
In this Letter we estimate the production of light elements including boron isotopes via the $\nu$-process in an SN Ic and find that SNe Ic can be viable $^{11}$B producers as well as SNe II. 
The resulting number abundance ratio of B isotopes from both the $\nu$-process and spallations  turns out to be $^{11}{\rm B}/^{10}{\rm B} = 3.7$ (for standard $T_{\nu_{\mu,\tau}}$) $- 4.3$ (for high $T_{\nu_{\mu,\tau}}$). 
It is reasonable to assume that the radii of neutrino spheres in SNe Ic are smaller than those in SNe II because of the compactness of SN Ic progenitors, leading to the higher temperatures of emitted neutrinos and anti-neutrinos.
Higher  $T_{\nu_{\mu \tau}}$ results in higher LiBeB yields through neutral current reactions in the $\nu$-process, which raises the significance of potential roles of SNe Ic in the light element production.
The $^{11}$B/$^{10}$B ratios inferred from current SN Ic models seem to agree well with solar values. However, the frequency of SNe Ic, in particular with high explosion energy as assumed here, is quite low and borons in meteorites should be dominated by those originated from SNe II and Galactic CRs. 
In fact, according to \citet{Cappellaro99} the estimated frequency ratio of SNe Ib/c to SNe II in local spiral galaxies falls in the range of 0.41 to 0.058 and the neutrino-processed $^{11}$B yield from each SN Ic is at most comparable to the contribution from SN II 
(averaged value per supernova $\sim 8.3 \times 10^{-7} \Msun$ from \citet{Rauscher02}; see also the case of SN 1987A model with similar parameter set in \citet{Yoshida08}). 
These facts lead to a conclusion that SNe Ic do not dominate the bulk of present $^{11}$B in the Galaxy.
Contribution from SNe Ic to the light element production in metal-deficient stars might be outstanding because an SN Ic progenitor is surrounded by its wind material and the light elements produced in the explosion are likely to be inherited directly by next generations of stars, which would show high BeB abundances overlying small amounts of BeB from the other processes. An extraordinary Be-rich halo star HD106038 \citep{Smiljanic08} could be a candidate star exhibiting such a process.

All previous Galactic chemical evolution (GCE) models included four different source components of the light element productions from primordial nucleosynthesis, AGB star and nova nucleosynthesis, SNe II $\nu$-process nucleosynthesis, and Galactic CR spallations and fusion \citep[][ and references therein]{Ryan01}.
Based on the GCE studies especially of the B isotope ratio, \citet{Yoshida05} discussed that the energy spectra of supernova neutrinos can be constrained through the avoidance of an overproduction of the $^{11}$B abundance \citep{Woosley95}.
Our present result in this Letter indicates that the GCE model should include the fifth astrophysical component of SNe Ic where both $\nu$-process and CNO CR spallation contribute remarkably to the production of $^7$Li and $^{11}$B.
The GCE studies including both contributions from SNe II and SNe Ic are highly desirable.

Neutrino oscillations change neutrino flavors in stellar materials.
It has been shown that light element yields in SNe II increase when the 
Mikheyev-Smirnov-Wolfstein (MSW) effect is taken into account 
\citep{Yoshida06a,Yoshida06b,Yoshida08}.
The contribution from charged-current $\nu$-process reactions becomes
more effective in the He layer because the spectra of $\nu_e$ and 
$\bar{\nu}_e$ shift to higher energy owing to the MSW effect. 
In the case of SNe Ic, however, the increase in the light element yields
by the MSW effect is expected to be small for the lack of He layer.
This relates to the locations of two resonances of the MSW effect.
Large flavor exchange occurs between $\nu_e$ ($\bar{\nu}_e$) and
$\nu_{\mu,\tau}$ ($\bar{\nu}_{\mu,\tau}$) at the higher density resonance
in the case of normal (inverted) mass hierarchy 
if $\sin^22\theta_{13} \ga 10^{-3}$.
The density of the higher resonance for a neutrino with energy $\ga 20$ MeV is smaller than 1,600 g cm$^{-3}$, 
corresponding to $M_r \ga 13.5 M_\odot$ in the mass coordinate.
Therefore, the $\nu$-process only in the outermost region in our SN Ic model will be affected
slightly by the MSW effect.

Recently, neutrino oscillations with neutrino self-interactions have been investigated \citep[e.g.][]{Duan06,Fogli07}.
The flavor change by this effect occurs in deep region of SN ejecta ($\sim 10^7$ cm) and the abundances of light element are expected to increase in entire region of the SN Ic.
This is because high-energy part of the neutrino spectra plays major role in the $\nu$-process nucleosynthesis where the neutrino self-interactions increase the contribution of charge current interaction in addition to the invariant neutral current interaction in the standard temperature hierarchy of supernova neutrinos which we set here $T_{\nu_e} < T_{\bar{\nu}_e} < T_{\nu_{\mu,\tau}}$.  Details of the effects of neutrino
self-interactions which depend on the initial spectra, luminosity,
density profile, and mixing angles of neutrinos will be discussed elsewhere.

The combined mechanism of light element production described here, i.e. the $\nu$-process and spallations, may also apply to the case of Type Ib supernovae where progenitor stars without H-rich envelope explode.
The outermost layers of SNe Ib is composed of He so that the $\alpha$-fusion reaction to produce lithium isotopes becomes effective.
Furthermore, \citet{Meynet02} suggested that the He layers of very metal-poor stars contain primary nitrogen which was produced during the He-burning phase by the CNO cycle in the H-burning shell stimulated by rotationally-induced diffusion of carbon into the H-burning shell. 
This strong thermonuclear reaction
may enhance the light element production via spallations because 
the compact progenitor resulted from strong mass loss leads to more effective acceleration of ejecta by shock wave
and the cross sections of spallation reactions involving nitrogen show high peak values at relatively low energy thresholds.
Observationally some low-metallicity halo stars were found to have high abundances of $^6$Li or $^9$Be beyond the theoretical predictions, which engages our interest in SNe Ib even if they might not accelerate their envelopes effectively because their progenitors are not so compact as those of SNe Ic.
A variety of progenitor models and parameters of SN explosion will be considered in the forthcoming paper, including their contributions to the Galactic chemical evolution.

This work has been partially supported by  the Grants-in-Aid for  Scientific Research (20105004, 20244035, 20540284, 21018004) of the Ministry of Education, Science, Culture, and Sports in Japan.

\end{document}